\documentclass[12pt]{article}

\usepackage{amssymb}
\usepackage{color}
\usepackage{epsfig,amssymb,amsfonts,amsmath,graphicx,dsfont,cite,xfrac}
\usepackage{authblk}
\usepackage{subfigure}
\definecolor{mygray}{gray}{0.5}

\usepackage{cite}
\usepackage[colorlinks=true,linkcolor=blue,citecolor=red]{hyperref}
\newcommand{\be}{\begin{equation}}
\newcommand{\ee}{\end{equation}}
\newcommand{\bea}{\begin{eqnarray}}
\newcommand{\eea}{\end{eqnarray}}

\title{Interplay between Riccati, Ermakov and Schr\"odinger  equations to produce complex-valued potentials with real energy spectrum}

\author{Zurika Blanco-Garcia}
\author{Oscar Rosas-Ortiz\thanks{Corresponding author. E-mail: orosas@fis.cinvestav.mx (O. Rosas-Ortiz).}}
\author{Kevin Zelaya}

\affil[]{\footnotesize Physics Department, Cinvestav, AP 14-740, 07000
M\'exico City, Mexico}

\date{}
\begin{document}

\maketitle

\begin{abstract}
Nonlinear Riccati and Ermakov equations are combined to pair the energy spectrum of two different quantum systems via the Darboux method. One of the systems is assumed Hermitian, exactly solvable, with discrete energies in its spectrum. The other system is characterized by a complex-valued potential that inherits all the  energies of the former one, and includes an additional real eigenvalue in its discrete spectrum. If such eigenvalue coincides with any discrete energy (or it is located between two discrete energies) of the initial system, its presence produces no singularities in the complex-valued potential. Non-Hermitian systems with spectrum that includes all the energies of either Morse or trigonometric P\"oschl-Teller potentials are introduced as concrete examples.

\end{abstract}

%%%%%%%%%%%%%%%%%%%%%%%%%%%%%%%%%%%%%%%%%%%%%%%

%---------------------------------------> Section
\section{Introduction}

The search for new integrable models in quantum mechanics has been the subject of intense activity during the last decades. The trend was firmly stimulated by the Witten formulation of supersymmetry published in 1981, where a quantum mechanical `toy model' was introduced as the simplest example of what occurs in quantum field theories \cite{Wit81}. The model evolved successfully into a thriving discipline that is nowadays known as supersymmetric quantum mechanics (Susy QM) \cite{Bag00,Coo01}. Sustained by the factorization method \cite{Mie04}, the supersymmetric approach is basically algebraic \cite{Mie84,And84a,And84b,And84c} and permits the pairing between the spectrum of a given (well known) Hamiltonian to the spectrum of a second (generally unknown) Hamiltonian. In position-representation, such pairing is ruled by a transformation introduced by Darboux in 1882 \cite{Dar82}. The latter implies that the involved potentials differ by an additive term, which in turn is the derivative of a function that solves a Riccati equation. Surprisingly, the origin of what later became known as Riccati's equation can be traced back to 1694, closely related to Bessel's equation, although the work of Riccati was published in 1724 \cite{Inc56,Wat66}. Besides Susy QM and soliton theory (where the Darboux transformation finds applications other than supersymmetry \cite{Rog02}), the presence of the Riccati equation is unavoidable in a lot of branches of physics and mathematics \cite{Sch18}. Its features in the complex domain \cite{Hil97} are the source of new challenges in controlling the time-evolution of quantum wave packets \cite{Cas13,Cru15,Cru16} as well as the paraxial propagation of structured light in optical media with quadratic index profile \cite{Cru17}. The complex version of the Riccati equation is also useful to strengthen the systematic search for non-Hermitian quantum systems with real spectrum \cite{Ros15}.

On the other hand, the stationary form of Schr\"odinger equation for one-dimensional systems is also connected with a nonlinear second-order differential equation introduced by Ermakov in 1880 \cite{Erm80}, and revisited by Milne fifty years after \cite{Mil30}. As in the Darboux approach, the Ermakov method implies that solving either of the two equations for any eigenvalue will provide a solution for the other. However, the Ermakov equation is usually linked to the time-evolution of systems like the isotonic oscillator \cite{Pin50,Lew67}, which is isochronous to the harmonic one in the classical picture \cite{Cha05} and isospectral to it in the quantum case \cite{Aso07}. Arnold \cite{Gue13,Gue15} and point \cite{Gun17,Car18} transformations facilitate the study of such systems. Nevertheless, the association of Ermakov with Schr\"odinger in spatial coordinates is rarely reported in the literature. Besides the monograph \cite{Sch18}, some exceptions can be found in \cite{Cru17,Ros15,Zel16,Jai17,Ros18,Kau95,Kau01,Ber03,Ber07,Kau00,Mos00}. Quite remarkably, only recently such relationship has been exploited to obtain integrable quantum models with real energy spectrum but described by non-Hermitian Hamiltonians \cite{Ros15}. A generalization of the oscillation theorem applies to study the zeros of the real and imaginary parts of the corresponding eigenfunctions \cite{Jai17}, and the introduction of bi-orthogonal bases permits to work with these systems in much the same way as for the Hermitian ones \cite{Ros18}. The non-Hermitian models so constructed are either PT-symmetric \cite{Ben05} or not. The main purpose of this work is to increase the number of solvable models in this direction and to bring to light the repercussion of combining nonlinear Riccati and Ermakov equations in the construction of new integrable models in quantum mechanics.

The paper is organized as follows: In section~\ref{basic} we develop the Darboux procedure to pair the spectrum of two different systems by including interplays between Schr\"odinger and Riccati, complex-valued Riccati and Ermakov, and between Ermakov and Schr\"odinger equations. In section~\ref{examples} we present the construction of new complex-valued potentials modeling non-Hermitian systems with the spectrum of either Morse or trigonometric P\"oschl-Teller systems; an additional real eigenvalue is included in each case. In section~\ref{quest} we discuss additional profiles of our method and show that the conventional one-step Darboux approach (producing new integrable Hermitian models) is recovered as a particular case. Of special interest, we also show that the construction of regular complex-valued potentials is viable even if the Darboux transformation is performed with the solution associated with an excited energy of the initial problem. Such a feature is not possible in the conventional one-step Darboux approaches since the validity of the oscillation theorem forbids the construction of potentials with no singularities if excited energies are used \cite{Sam99}. The paper ends with some concluding remarks.

%---------------------------------------> Section
\section{Basic formalism}
\label{basic}

Within the Darboux approach \cite{Dar82}, the second-order differential equation
\be
-\varphi''+ V_0(x) \varphi = E \varphi
\label{sturm1}
\ee
can be transformed into a new one
\be
-\psi'' + V(x) \psi = E \psi
\label{sturm2}
\ee
through the functions
\be
V(x) = V_0(x) + 2 \beta', \quad \psi= \varphi' + \beta \varphi,
\label{darboux}
\ee
where $f'$ denotes the derivative of $f$ with respect to $x$ and $\beta$ is defined by the nonlinear Riccati equation
\be
-\beta' +\beta^2 =V_0(x) -\epsilon.
\label{riccati1}
\ee
Using
\be
\beta = -(\ln u)',
\label{beta}
\ee
the latter equation is linearized
\be
-u'' +V_0(x) u = \epsilon u.
\label{u}
\ee
In other words, $u$ is a solution of (\ref{sturm1}) with $E=\epsilon$. The link (\ref{beta}) between the nonlinear first order differential equation (\ref{riccati1}) and the linear second order one (\ref{u}) is well known in the literature \cite{Inc56,Hil97}. This can be also written as the first order differential equation $u'+\beta u=0$. Thus, provided a solution of (\ref{riccati1}), the function
\be
u(x) = \mbox{const} \times \exp \left[-\int^x \beta(y) dy\right]
\label{bsol}
\ee
solves (\ref{u}). One can go a step further by adding $2 \beta'$ to both sides of (\ref{riccati1}), then $\beta' +\beta^2 =V(x) -\epsilon$, where (\ref{darboux}) has been used. Instead of (\ref{beta}) we now make $\beta = (\ln \psi_{\epsilon} )'$, so the latter Riccati equation is also linearized: $-\psi_{\epsilon}'' + V(x) \psi_{\epsilon} = \epsilon \psi_{\epsilon}$. The straightforward calculation shows that $\psi_{\epsilon} \propto u^{-1}$. In summary, the covariance (\ref{sturm1})-(\ref{sturm2}) revealed by the Darboux transformation (\ref{darboux}) allows new integrable models if the spectrum properties of either $V_0(x)$ or $V(x)$ are already known.

%---------------------------------------> Subsection
\subsection{Schr\"odinger-Riccati interplay}

In quantum mechanics the eigenvalue equation (\ref{sturm1}) is defined by the energy observable of a particle with one degree of freedom in the stationary case. For real-valued measurable functions $V_0(x)$, the eigenvalues $E$ are real and the Sturm-Liouville theory applies on the eigenfunctions $\varphi$ representing bound states. As indicated above, the method introduced by Darboux is very useful to construct new solvable potentials $V(x)$. Indeed, if the solutions of the Schr\"odinger equation (\ref{sturm1}) are already known, the energy spectrum of $V_0(x)$ is entirely inherited to $V(x)$ for the appropriate solution of the Riccati equation (\ref{riccati1}) \cite{Mie84,And84a}. Besides, if $\psi_{\epsilon} \propto u^{-1}$ is physically admissible, the spectrum of $V(x)$ includes also the eigenvalue $\epsilon$ \cite{Mie84}. In algebraic form, the Darboux transformation (\ref{darboux}) results from the factorization of the Hamiltonians defined by $V_0(x)$ and $V(x)$ \cite{Mie84,And84b}, and represents the kernel of Susy QM \cite{Bag00,Coo01,Mie04}. Typically, real-valued functions $\beta$ are used to construct new Sturm-Liouville integrable equations (\ref{sturm2}). In such case $\epsilon \leq E_0$, where $E_0$ is the ground state energy if $V_0(x)$ has discrete spectrum and the low bound of the continuum spectrum when the bound states are absent \cite{Sam99}. Therefore, the new potential $V(x)$ is real-valued and defines a Hermitian Hamiltonian. However, this is not the only option permitted by the method \cite{Mie04}. Indeed, interesting non-Hermitian models arise if $\beta$ is allowed to be complex-valued since the entire spectrum of $V_0(x)$ is still inherited to $V(x)$  \cite{Can98,And99,Bag01,Ros03,Ros07,Fer08}.

In the following we show that the relationship between $\beta$ and $u$ is not uniquely determined by the logarithmic derivative (\ref{beta}). Our approach brings to light some nonlinear connections between the systems associated with $V_0(x)$ and those defined by $V(x)$ that are hidden in the conventional way of dealing with Eqs.~(\ref{sturm1})-(\ref{darboux}).

%---------------------------------------> Subsection
\subsection{Riccati-Ermakov interplay}

Let us look for complex-valued solutions $\beta = \beta_R + i \beta_I$ of the Riccati equation (\ref{riccati1}) with $\epsilon \in \mathbb R$. Here $\beta_R$ and $\beta_I$ are real-valued functions to be determined. Following \cite{Ros15} we see that the real and imaginary parts of Eq.~(\ref{riccati1}) lead to the coupled system
\be
-\beta_R' + \beta_R^2 - \beta_I^2 + \epsilon -V_0(x)=0,
\label{rica1}
\ee
\be
-\beta_I' + 2 \beta_I \beta_R =0.
\label{rica2}
\ee
Rewriting (\ref{rica2}) as $(\ln \beta_I)' = 2 \beta_R$, one immediately notice that $\beta_R =-(\ln \alpha)'$ gives rise to the expression $\beta_I = \frac{\lambda}{\alpha^2}$, where $\lambda \in \mathbb R$ is a constant of integration and $\alpha$ is a function to be determined. To avoid singularities in $\beta_R$ and $\beta_I$ we shall consider real-valued $\alpha$-functions with no zeros in $\mbox{Dom} V_0 \subseteq \mathbb R$. Notice however that $\alpha$ is permitted to be purely imaginary also. The introduction of $\beta_R$ and $\beta_I$ into (\ref{rica1}) gives the nonlinear second order differential equation
\be
\alpha'' = \left[ V_0(x) - \epsilon \right] \alpha + \frac{\lambda^2}{\alpha^3},
\label{ermakov}
\ee
which is named after Ermakov \cite{Erm80}. Provided a solution of (\ref{ermakov}), the complex-valued function we are looking for is 
\be
\beta_{\lambda} = -(\ln \alpha)' + i \frac{\lambda}{\alpha^2}.
\label{betac}
\ee
The sub-label of $\beta$ indicates that it is separable into $\beta_R$ and $\beta_I$ only when $\lambda \neq 0$. Indeed, $\lambda=0$ reduces (\ref{ermakov}) to the linear equation (\ref{u}), and brings (\ref{betac}) to its usual form (\ref{beta}). 

%---------------------------------------> Subsection
\subsection{Ermakov-Schr\"odinger interplay}
\label{E-S}

Hereafter we take $\lambda \neq 0$ and assume that $u_p$ is a particular solution of (\ref{u}). To construct $\alpha$ we follow \cite{Erm80} and eliminate $V_0(x) - \epsilon$ from (\ref{ermakov}) and (\ref{u}). It yields the equation
\be
W'(u_p,\alpha) = \frac{\lambda^2 u_p }{\alpha^3},
\label{wprim}
\ee
where $W(u_p,\alpha) = u_p \alpha' - u_p' \alpha$ is the Wronskian of $u_p$ and $\alpha$. Multiplying both sides of (\ref{wprim}) by $2W(u_p,\alpha)$, and making a simple integration we have
\be
J= W^2(u_p,\alpha) + \left( \frac{ \lambda u_p}{\alpha} \right)^2,
\label{J}
\ee
where $J$ is an integration constant\footnote{The structure of $J$ coincides with the invariant $I$ found by Lewis for the time-evolution of the isotonic oscillator \cite{Lew67}. Indeed, after the identification $x \rightarrow t$, one may be tempted to find a `physical' meaning for $J$ (see for instance \cite{Kau00} and the discussion on the matter offered in \cite{Mos00}). However, although $J$ plays a central role in our approach, we shall take it just as it is: an integration constant.}. As we have no way to fix the value of $J$ a priori, let us consider $J=0$ and $J\neq 0$ separately. In the former case Eq.~(\ref{J}) is reduced to 
\be
W(u_p, \alpha_0) = \pm i \frac{\lambda u_p}{\alpha_0}.
\label{W}
\ee
The sub-label of $\alpha$ means $J=0$. Solving (\ref{W}) for $u_p$ we obtain (remember $\lambda \neq 0$):
\be
u_p(x) = c_0 \alpha_0(x) \exp \left[ \mp i \lambda \int^x \alpha_0^{-2} (y) dy \right],
\label{ufunction}
\ee
with $c_0$ an integration constant. This result is consistent with \cite{Kau00,Mos00}. To get new insights on the properties of Eq.~(\ref{W}) let us rewrite it in simpler form
\be
\frac{d}{dx} \left( \frac{\alpha_0}{u_p} \right)^2 = \pm i \frac{2\lambda}{u_p^2},
\label{W2}
\ee
where we have used  
\be
W(u,\alpha) = u^2 \left( \frac{\alpha}{u} \right)'.
\label{newW}
\ee
After integrating (\ref{W2}) we arrive at
\be
\alpha_0^2 (x) = \pm i 2 \lambda \left[ u_p(x) q(x) \right] u_p(x) + c_{\alpha} u_p^2(x),
\label{apart}
\ee
with $c_{\alpha}$ an integration constant, and
\be
q(x) = \int^x u_p^{-2}(y) dy.
\label{q}
\ee
Equation (\ref{apart}) reveals that $\alpha_0$ is not only connected with the particular solution $u_p$, but it is indeed associated with a fundamental pair of solutions of (\ref{u}). For if the wronskian $W(u_p,v)=\omega_0$ is a constant different from zero, it may be proven that $v(x) = \omega_0 u_p(x) q(x)$ is a second linearly independent solution of (\ref{u}) \cite{Arf01}. Therefore, $\alpha_0$ is expressed in terms of the basis $u_p$ and  $v$ as follows
\be
\alpha_0(x)= \pm \left[ \pm i \frac{2\lambda}{\omega_0} v(x) u_p(x) + c_{\alpha} u_p^2(x)
\right]^{1/2}.
\label{apart2}
\ee

On the other hand, for $J\neq 0$ it is convenient to rewrite Eq.~(\ref{J}) as 
\be
\pm \frac{2}{u_p^2} \left[ J \left( \frac{\alpha}{u_p} \right)^2 - \lambda^2 \right]^{1/2} = \frac{d}{dx} \left( \frac{\alpha}{u_p}
\right)^2,
\ee
where we have used (\ref{newW}). The proper rearrangements and a simple integration produce
\be
 J q(x)+ I_0  = \pm \left[ J \left( \frac{\alpha}{u_p} \right)^2 - \lambda^2 \right]^{1/2},
 \label{root}
\ee
with $I_0$ a new integration constant. Now we can solve (\ref{root}) for $\alpha$ to get
\be
\alpha(x) = \pm \left[ av^2(x) + b v(x) u_p(x) + c u_p^2(x) \right]^{1/2}.
\label{alpha}
\ee
A simple calculation shows that the set 
\be
a= \frac{J}{\omega_0^2}, \quad b= 2 \frac{I_0}{\omega_0}, \quad c= \frac{\lambda^2 +I_0^2}{J},
\label{const}
\ee
satisfies $4ac- b^2 = 4 (\lambda/\omega_0)^2$ \cite{Ros15}.  Although the structure of (\ref{apart2}) and (\ref{alpha}) is quite similar, some caution is necessary to recover $\alpha_0$ from $\alpha$ since $c$ may be ill defined for $J=0$. To get some insight on the matter consider the limit $J \rightarrow 0$ in (\ref{root}); this yields $I_0 \rightarrow \pm i \lambda$. Consequently $b \rightarrow \pm i \frac{2\lambda}{\omega_0}$, which is the coefficient of $v(x)u_p(x)$ in (\ref{apart2}), and $a \rightarrow 0$. Therefore, if $c \rightarrow \mbox{const}$ as $J \rightarrow 0$, we can take $c \rightarrow c_{\alpha}$ to get $\alpha \rightarrow \alpha_0$.

For simplicity, in the sequel we take $a,b,c \in \mathbb R$ (equivalently, the integration constants $J \neq 0$, $I_0$, as well as $\omega_0$, are real). Since $\lambda \neq 0$ we immediately see that $ac > (b/2)^2$. Thus, $a$ and $c$ will have the same sign (which is determined by $J\neq 0$) and both are different from zero. In turn, $b$ is permitted to be any real number. As indicated above, the $\alpha$-function can be purely imaginary in (\ref{betac}). This occurs if, for instance, $b=0$ and $J<0$. Although it is not necessary, we shall take $J>0$ to have real-valued $\alpha$-functions (for the examples discussed in the next sections the sign of $I_0$ does not affect such a condition). In addition, without loss of generality, we  shall take the root `$+$' of (\ref{alpha}) as the $\alpha$-function of our approach.

%---------------------------------------> Subsection
\subsection{Properties of the fundamental solutions}
\label{properties}

Provided $\alpha$, the new potential $V(x)$ is complex-valued and parameterized by $\lambda$,
\be
V_{\lambda}(x) = V_0(x) - 2(\ln \alpha)'' + i 2 \left(\frac{\lambda}{\alpha^2} \right)'.
\label{newpot}
\ee
To show that $\alpha$ is free of zeros in $\mbox{Dom} V_0$ let us suppose that $\alpha=0$ in any interval ${\cal I} \subset \mbox{Dom} V_0$, which may be of measure zero. From (\ref{alpha}) we see that this implies $v=\frac{1}{2a} ( -b \pm i \frac{2 \lambda}{\omega_0} ) u_p$ for $x \in {\cal I}$. In other words, $v$ should differ from $u_p$ by a multiplicative constant in ${\cal I}$. However, this is not possible since  $u_p$ and $v$ are linearly independent in $\mbox{Dom} V_0$. On the other hand, given a bound state $\psi$ of potential (\ref{newpot}), the conventional notions of probability density $\rho = \vert \psi \vert^2$ and probability current ${\cal J}= i \left( \psi \frac{\partial \psi^*}{\partial x}  - \frac{\partial \psi}{\partial x} \psi^* \right)$ lead to the continuity equation $\frac{\partial {\cal J}}{\partial x} + \frac{\partial \rho}{\partial t} = 2 \mbox{Im} V_{\lambda} (x)$ \cite{Zel16}. Integrating over $\mbox{Dom} V_0$ (at time $t_0$) we see that ${\cal J}$ is twice the area defined by $\mbox{Im} V_{\lambda}(x)$. If such area is reduced to zero then the total probability is conserved, although spatial variations of ${\cal J}$ may be not compensated by temporal variations of $\rho$ locally. In this context the  {\em condition of zero total area} \cite{Jai17},
\be
\int_{Dom V_0} \mbox{Im} V_{\lambda} (x) dx =  \left. \frac{2 \lambda}{\alpha^2} \right\vert_{Dom V_0} =0,
\label{zero}
\ee
ensures conservation of total probability. 

We would like to emphasize that one-dimension potentials featuring the parity-time (PT) symmetry represent a particular case of the applicability of (\ref{zero}). Such potentials are invariant under parity (P) and time-reversal (T) transformations in quantum mechanics \cite{Ben05}. The former corresponds to spatial reflection $p \rightarrow -p$, $x \rightarrow -x$, and the latter to $p \rightarrow -p$, $x \rightarrow x$, together with complex conjugation $i \rightarrow -i$. Thus, a necessary condition for PT-symmetry is $V(x)=V^*(-x)$, where ${}^*$ stands for complex conjugation. For initial potentials $V_0(x)$ such that $V_0(x)= V_0(-x)$, one can show that making $b=0$ in (\ref{alpha}) is sufficient to get $V_{\lambda}(x) = V^*_{\lambda}(-x)$; some examples are given in Section~\ref{examples}. In other words, PT-symmetry is a consequence of (\ref{zero}) in our approach, so this symmetry is not a necessary condition to get complex-valued potentials with real spectrum in general. Therefore, potentials $V_{\lambda}(x)$ that satisfy (\ref{zero}) can be addressed to represent open quantum systems with balanced gain (acceptor) and loss (donor) profile \cite{Ele17}, no matter if they are PT-symmetric or not.

In our model condition (\ref{zero}) is satisfied by using $\alpha$-functions that diverge at the edges of $\mbox{Dom} V_0$. In such case the complex-valued eigenfunctions $\psi(x)$ that are obtained from the transformation of bound states $\varphi(x)$ can be normalized in conventional form \cite{Ros15}. Then, $\vert \psi(x) \vert^2$ defines a finite area in any interval of $\mbox{Dom} V_0$ and the distribution of its maxima is quite similar to that of the conventional probability densities \cite{Jai17}. Moreover, it can be shown that the real and imaginary parts of $\psi(x)$ obey interlacing theorems that are very close to those satisfied by $\varphi(x)$ \cite{Jai17}. The latter permits a bi-orthogonal approach in which the spectral properties of $V_{\lambda}(x)$ can be studied in much the same way as in the real-valued case \cite{Ros18}. For practical purposes, instead of the conventional normalization, we shall use the bi-normalization introduced in \cite{Ros18} for the eigenfunctions $\psi(x)$ that are associated with discrete energies (the quantitative difference is short). In the sequel, all the eigenfunctions of $V_{\lambda}(x)$ will be written as $\psi^{\lambda}(x)$. Additional labels may be included to distinguish between bound and other kind of energy states.

To conclude this section let us introduce (\ref{betac}) into (\ref{bsol}), the result is a seed function $u$ that is now labeled by $\lambda$. As indicated above, the reciprocal of $u_{\lambda}$ gives $\psi_{\epsilon}^{\lambda} \propto u_{\lambda}^{-1}$, which is a solution of (\ref{sturm2}) for $E=\epsilon$. Using Eq.~2.172 of \cite{Gra}, after some simplifications, we obtain
\be
\psi^{\lambda}_{\epsilon} (x)= 
\frac{c_{\epsilon}}{\alpha^2(x)} \left[ \frac{1}{c} \left( \frac{\lambda}{\omega_0} -i \frac{b}{2}  \right) v(x) -i u_p(x)  \right],
\label{missing}
\ee
where the constant $c_{\epsilon} $ may be fixed by normalization. 

%---------------------------------------> Section
\section{Examples}
\label{examples}

The following examples consider potentials $V_0(x)$ that include discrete eigenvalues in their energy spectra. We shall focus on the discrete energies of the new potentials $V_{\lambda}(x)$; the study of scattering states, if they are included in the spectrum of $V_0(x)$, will be reported elsewhere. Two cases of $V_0(x)$ are analyzed: the Morse potential, which is defined over all the real line and allows discrete as well as scattering energies, and the trigonometric P\"oschl-Teller potential, which is finite in a concrete interval of $\mathbb R$, and admits discrete spectrum only. Our purpose is to illustrate that the method introduced in the previous sections works very well in any domain $\mbox{Dom}V_0 \subseteq \mathbb R$. 

%---------------------------------------> Subsection
\subsection{Morse potentials}

\begin{figure}[htb]
\centering
\includegraphics[width=0.5\textwidth]{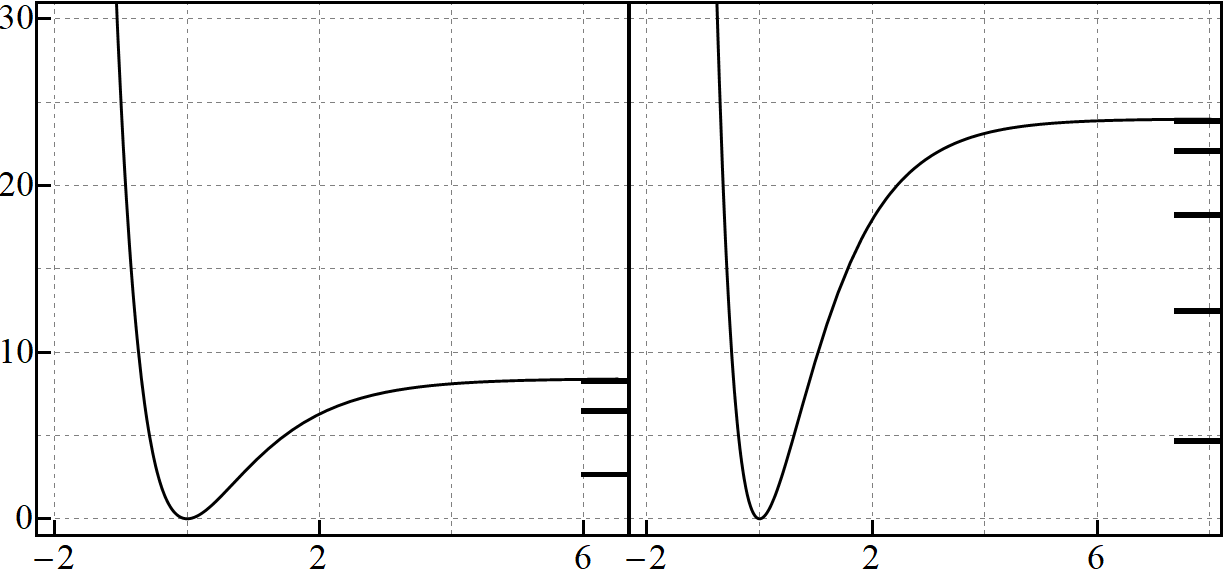}

\caption{\footnotesize 
Morse potential (\ref{morse1}) of depth (\ref{depth}) with $\gamma=1$, $\delta=0.4$,  $N=2$ (left) and $N=4$ (right). This kind of potentials permits the presence of $N+1$ bound states. In each case the allowed discrete energies (\ref{morsen}) are ticked on the right-vertical scale. 
}
\label{FigM1}
\end{figure}

The eigenvalue equation (\ref{sturm1}) for the Morse potential (see Fig.~\ref{FigM1}),
\be
V_0 (x)= \Gamma_0(1-e^{-\gamma x})^{2}, \quad  \Gamma_0, \gamma >0, \quad \mbox{Dom} V_0 = \mathbb R,
\label{morse1}
\ee
admits the fundamental set of solutions
\bea
u_p(y)=e^{-y/2} y^{\sigma} {}_{1}F_{1}\left( \sigma +\frac{1}{2} - d ; 1+2\sigma ; y \right),
\label{solM1}\\[1ex]
v( y )=e^{-y/2} y^{-\sigma} {}_{1}F_{1}\left( -\sigma +\frac{1}{2} - d ; 1-2\sigma ; y \right),
\label{solM2}
\eea

%%%%%%%%%%%%%
%%%%%%%%%%%

\noindent
where ${}_1F_1({\rm a},{\rm c};z)$ stands for the hypergeometric confluent function \cite{Olv10}, and
\be
y = 2 d e^{-\gamma x}, \quad d^{2}=\frac{\Gamma_{0}}{\gamma^2}, \quad \sigma =\frac{\Gamma_{0}^2-E}{\gamma^{2}}.
\ee
A straightforward calculation shows that $W(u_p,v) \equiv w_{0}=2 \sqrt{\Gamma_0 -E}$. To determine the bound states it is convenient to rewrite the potential depth $\Gamma_0$ as 
\be
\Gamma_0 =\gamma^2(N+\delta+1/2)^{2}, \quad N\in\mathbb{Z}^{+}\cup\{0\}, \quad 0<\delta<1.
\label{depth}
\ee
Potentials (\ref{morse1}) with depth (\ref{depth}) are depicted in Fig.~\ref{FigM1} for $N=2$ and $N=4$. 

\begin{figure}[htb]
\centering
\includegraphics[width=0.3\textwidth]{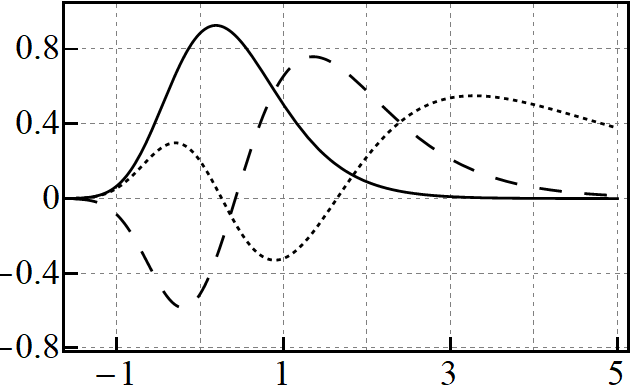}

\caption{\footnotesize 
The bound states (\ref{solMc1}) of the Morse potential depicted in Fig.~\ref{FigM1}~($N=2$) with  
$n=0$ (solid), $n=1$ (dashed), and $n=2$ (dotted).}
\label{FigM2}
\end{figure}

The set of discrete energies is therefore finite and defined by the expression
\be
E_{n}=\gamma^{2}\left[ (2n+1)(N+\delta+1/2) - (n+1/2)^2 \right], \quad n=0,1,\ldots, N.
\label{morsen}
\ee
The corresponding eigenfunctions are obtained from $u_p$ with $\sigma +\frac{1}{2} - d =n \in \mathbb{Z}^+ \cup \{ 0 \}$,
\be
 \varphi_n (y)= \mathcal{C}_{n} e^{-y/2} y^{\alpha_{n}} L_{n}^{(2\alpha_n) }(y), \quad n=0,1,\ldots, N,
\label{solMc1}
\ee
where $L_{n}^{(\alpha)}(z)$ stands for the associated Laguerre polynomials \cite{Olv10}, and
\be
\alpha_{n}=d-1/2-n, \quad \mathcal{C}_{n}^{2}=\frac{\gamma(2d-1-2n)n!}{\Gamma(2d-n)}.
\ee
Note that we have fixed $N=n_{max}=\lfloor d - 1/2 \rfloor$, with $\lfloor a \rfloor$ the floor function \cite{Olv10}. The bound states for $N=2$ are shown in Fig.~\ref{FigM2}.

%%%%%%%%%%%%%
%%%%%%%%%%%

\begin{figure}[htb]
\centering
\subfigure[$N=2$]{\includegraphics[width=0.3\textwidth]{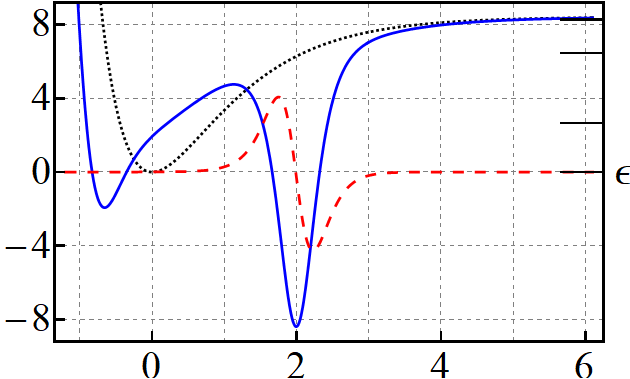}} 
\hspace{4ex}
\subfigure[$N=4$]{\includegraphics[width=0.3\textwidth]{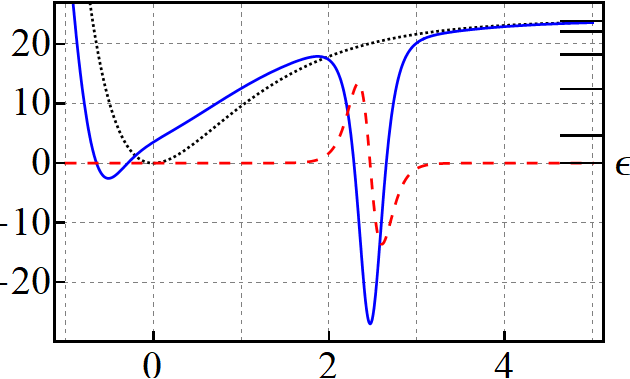}}

\caption{\footnotesize 
(Color online) Real (blue-solid) and imaginary (red-dashed) parts of $V_{\lambda}(x)$, which is generated from the Morse potentials (black-dotted) of Fig.~\ref{FigM1}. Here, $\lambda=J=I_0=1$, and $\epsilon=0$. The ticks on the right-vertical scale refer to the discrete energies (\ref{morsen2}). 
}
\label{FigM3}

\end{figure}

Now, to construct a complex-valued potential $V_{\lambda}(x)$ from the Morse family (\ref{morse1}) we may make $\epsilon < E_0 = \gamma^2 (N + \delta + 1/4)$. The discrete energy spectrum of such system is also finite, and includes $N+2$ eigenvalues
\be
E_0^{\lambda}= \epsilon, \quad E_{n+1}^{\lambda} =\gamma^{2}\left[ (2n+1)(N+\delta+1/2) - (n+1/2)^2 \right], \quad n=0,1,\ldots, N.
\label{morsen2}
\ee
The new potential is shown in Fig.~\ref{FigM3} for $\epsilon=0$ and two different values of $N$. Notice that $\mbox{Re} V_{\lambda}(x)$ is very close to the Morse potential at the edges of $\mathbb R$. Besides, such a function exhibits a very localized deformation that serves to host the additional energy $E^{\lambda}_0  =0$. In turn, $\mbox{Im} V_{\lambda}(x)$ satisfies the condition of zero total area (\ref{zero}). Clearly, these potentials are not eligible for PT-transformations.

%%%%%%%%%%%%%
%%%%%%%%%%%

\begin{figure}[htb]
\centering
\subfigure[$\psi^{\lambda}_0(x)$]{\includegraphics[width=0.3\textwidth]{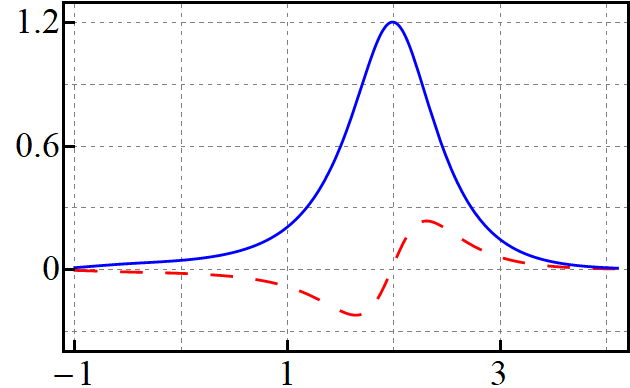}} 
\hspace{1ex}
\subfigure[$\psi^{\lambda}_1(x)$]{\includegraphics[width=0.3\textwidth]{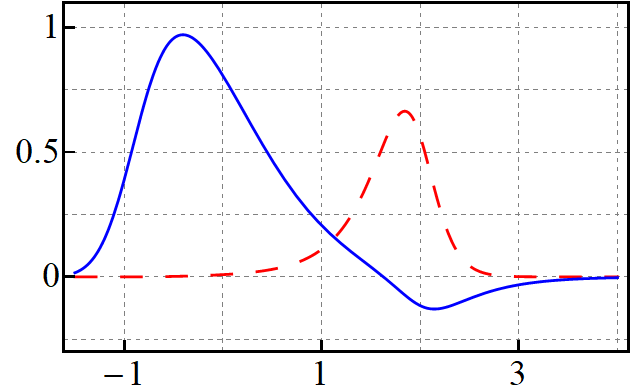}}
\hspace{1ex}
\subfigure[$\psi^{\lambda}_2(x)$]{\includegraphics[width=0.3\textwidth]{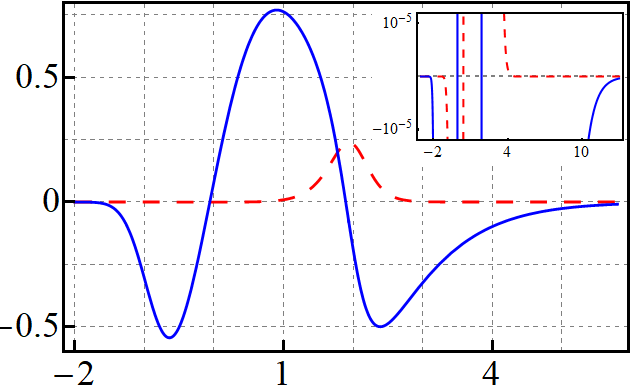}}

\vskip1ex
\subfigure[$\psi^{\lambda}_0(x)$]{\includegraphics[width=0.3\textwidth]{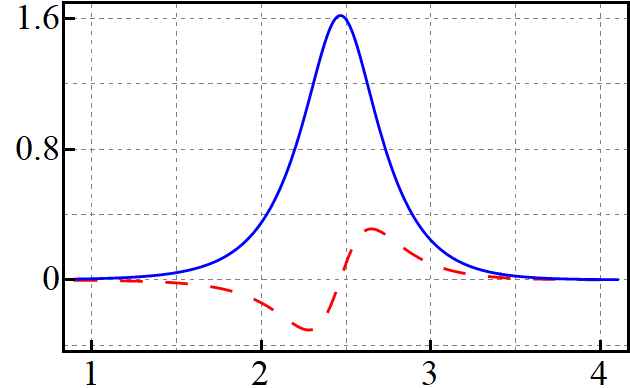}} 
\hspace{1ex}
\subfigure[$\psi^{\lambda}_1(x)$]{\includegraphics[width=0.3\textwidth]{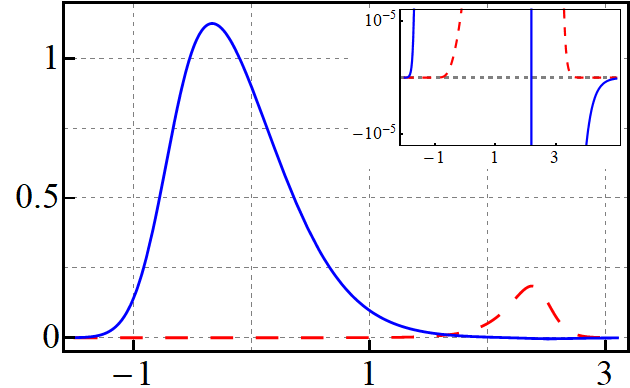}}
\hspace{1ex}
\subfigure[$\psi^{\lambda}_2(x)$]{\includegraphics[width=0.3\textwidth]{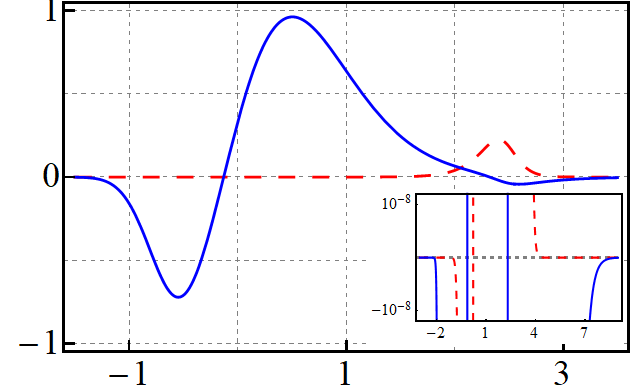}}

\caption{\footnotesize 
(Color online) Real (blue-solid) and imaginary (red-dashed) parts of the eigenfunctions $\psi_n(x)$ associated with potentials of Fig.~\ref{FigM3} for the indicated values of $n$, with $N=2$ (upper row), $N=4$ (lower row). For the sake of visibility, the plots of the imaginary parts in (b), (e), and (f) have been rescaled by a factor of 10, 100 and 10, respectively. The insets in (c), (e), and (f) show the interlacing of zeros for the corresponding functions.
}
\label{FigM4}

\end{figure}

The related bi-normalized complex-valued eigenfunctions $\psi^{\lambda}_n(x)$ are depicted in Fig.~\ref{FigM4}. These new functions do not form an orthogonal set but the zeros of their real and imaginary parts are interlaced according to the theorems reported in \cite{Jai17}. Namely, between two zeros of $\mbox{Re} \, \psi^{\lambda}_{n+1}$ there is always a zero of $\mbox{Im} \,\psi^{\lambda}_{n+1}$, with $n=0,1,\ldots, N.$

%%%%%%%%%%%%%
%%%%%%%%%%%

\noindent
$\bullet$ {\bf Oscillator potentials.} Departing from the `mathematical' oscillator $V_0(x) =x^2$, one recovers the family of complex-valued oscillators $V_{\lambda}(x)$ introduced in \cite{Ros15} and studied in \cite{Zel16,Jai17,Ros18}. Since the Morse potential (\ref{morse1}) converges  to $x^2$, it may be shown that the complex-valued potentials derived in this section converge to the oscillators reported in \cite{Ros15} at the appropriate limit.
 
%---------------------------------------> Subsection
\subsection{Trigonometric P\"oschl-Teller potentials}

%%%%%%%%%%%%%
\begin{figure}[htb]
\centering
\includegraphics[width=0.5\textwidth]{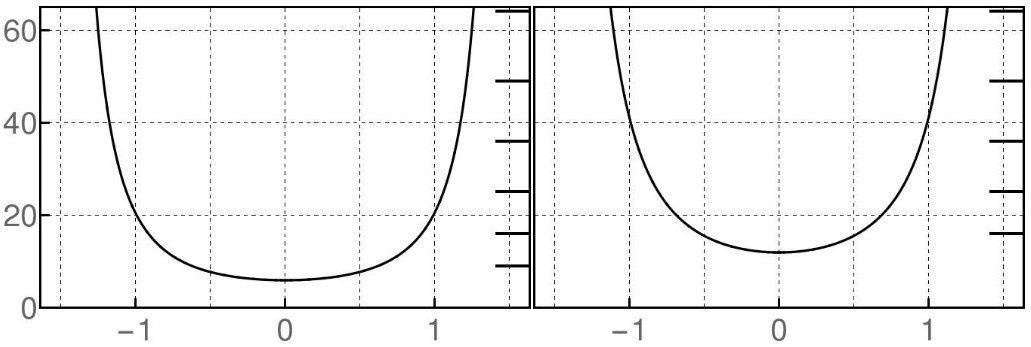}

\caption{\footnotesize 
Trigonometric P\"oschl-Teller potential (\ref{PT}) for $U_0=1$ with $r=3$ (left) and $r=4$ (right). In each case the ticks on the right-vertical scale refer to the discrete energies (\ref{enerPT}).
}
\label{FigPT1}
\end{figure}
%%%%%%%%%%%

\noindent
The eigenvalue equation (\ref{sturm1}) for the trigonometric P\"oschl-Teller potential (see Fig.~\ref{FigPT1}),
\be
V_0(x)= \left\{
\begin{array}{ll}
U_0^2\frac{r(r-1)}{\cos^2( U_0 x)}  & x \in \left( -\frac{\pi}{2 U_0}, \frac{\pi}{2 U_0} \right) \\[1.5ex]
+\infty & \mbox{otherwise}
\end{array}
\right. ,
\quad r>1, \quad U_0 >0,
\label{PT}
\ee
is solved by the linearly independent functions 
\bea
u_p (x)=\cos^r (U_0 x) \, {}_2F_1\left[  {\rm a}, {\rm b}, {\rm c}; \sin^2(U_0 x) \right],
\label{solPT1}\\[2ex]
v(x)=\cos^r (U_0 x) \sin(U_0 x) \, {}_2F_1\left[  {\rm a} + \frac12, {\rm b} + \frac12, {\rm c} +1; \sin^2(U_0 x) \right],
\label{solPT2}
\eea
where ${}_2F_1({\rm a}, {\rm b}, {\rm c};z)$ is the hypergeometric function \cite{Olv10}, with
\be
{\rm a} = \frac{1}{2} \left( r + \frac{\sqrt{E}}{U_0} \right), \quad {\rm b} = \frac{1}{2} \left( r -\frac{\sqrt{E}}{U_0} \right), \quad {\rm c} = \frac12.
\ee

%%%%%%%%%%%%%
\begin{figure}[htb]
\centering
\includegraphics[width=0.3\textwidth]{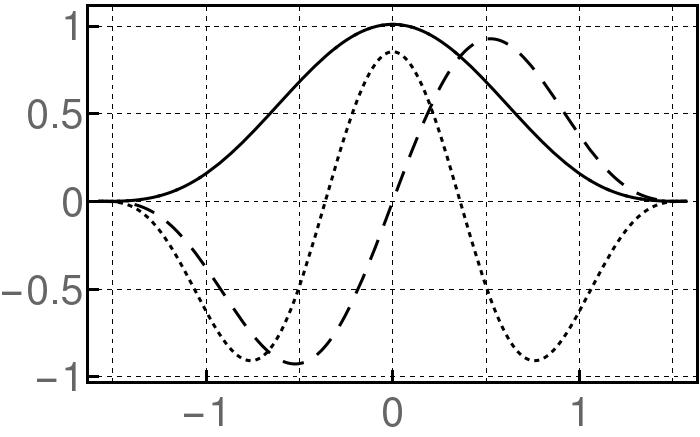}

\caption{\footnotesize 
First eigenfunctions (\ref{even})-(\ref{odd}) of the trigonometric P\"oschl-Teller potential depicted in Fig.~\ref{FigPT1} ($r=3$) with  $n=0$ (solid), $n=1$ (dotted), and $n=3$ (dashed). 
}
\label{FigPT2}
\end{figure}
%%%%%%%%%%%

\noindent
The hypergeometric functions in (\ref{solPT1}) and (\ref{solPT2}) are linearly independent at the regular singularity $z=0$ of the hypergeometric equation. The one in (\ref{solPT1}) is the only which is analytic at $z=0$. For the function defining $v(x)$ in (\ref{solPT2}), $z=0$ is a branch point. Using the Wronskian of such functions (c.f.  Eq.~15.10.3 of \cite{Olv10}) it can be proven that $W(u_p,v) \equiv \omega_0 = U_0$. Then, the physical solutions are obtained by studying the behavior of ${}_2F_1({\rm a}, {\rm b}, {\rm c};z)$ with argument unity. As ${\rm c} - {\rm a} - {\rm b} <0$, we use Eq.~15.8.4 of \cite{Olv10} and realize that $u_p(x)$ satisfies the boundary conditions if either ${\rm a}$ or ${\rm b}$ is a negative integer. The latter leads to the even bound states (see Fig.~\ref{FigPT2}),
\be
\varphi_{2n}(x)= \nu_{2n} \cos^r (U_0 x) \, {}_2F_1\left[ -n, r+n, \tfrac12; \sin^2(U_0 x) \right].
\label{even}
\ee
In turn, if either ${\rm a} +\tfrac12$ or  ${\rm b} +\tfrac12$ is a negative integer, the function $v(x)$  gives the odd bound states (see Fig.~\ref{FigPT2}),
\be
\varphi_{2n+1}(x)= \nu_{2n+1} \cos^r (U_0 x) \sin(U_0 x)  \, {}_2F_1\left[ -n, r+n+1, \tfrac32; \sin^2(U_0 x) \right].
\label{odd}
\ee
The symbol $\nu_n$ in the above expressions stands for the normalization constant. The energy spectrum of trigonometric P\"oschl-Teller potential (\ref{PT}) is therefore defined by the discrete set
\be
E_n= U_0^2(n+r)^2, \quad n=0,1,2,\ldots
\label{enerPT}
\ee

%%%%%%%%%%%%%
\begin{figure}[htb]
\centering
\subfigure[$r=3$]{\includegraphics[width=0.3\textwidth]{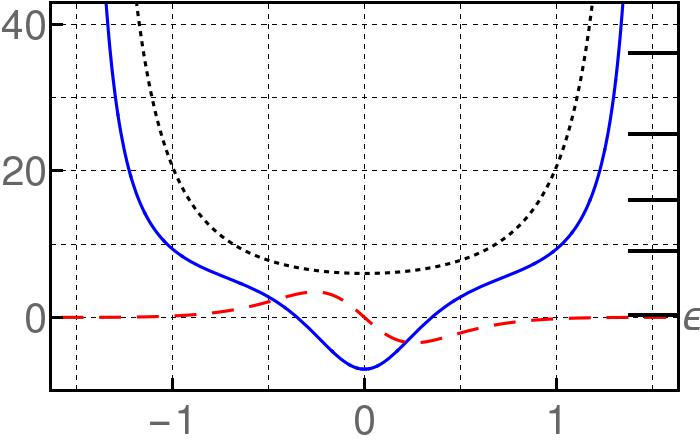}} 
\hspace{5ex}
\subfigure[$r=4$]{\includegraphics[width=0.3\textwidth]{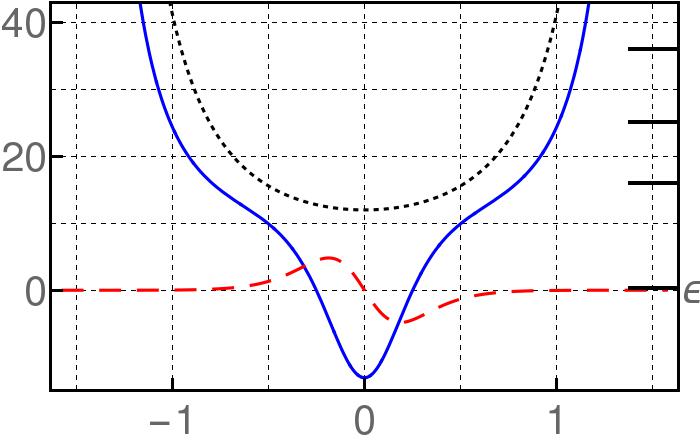}}

\caption{\footnotesize 
(Color online) Real (blue-solid) and imaginary (red-dashed) parts of the PT-symmetric version of potentials depicted in Fig.~\ref{FigPT1} (black-dotted). Here $J=\pi/4$, $I_0=0$, and $\lambda=\sqrt{\pi/4}$. The ticks on the right-vertical scale refer to energies (\ref{enerPT2}), with $\epsilon=1/4$.
}
\label{FigPT3}

\end{figure}
%%%%%%%%%%%

In the present case we use  $\epsilon <E_0 =U_0^2r^2$ to generate complex-valued potentials $V_{\lambda}(x)$ that include the set (\ref{enerPT}) in their energy spectrum. That is, the energy eigenvalues of $V_{\lambda}(x)$ are defined by the denumerable set
\be
E^{\lambda}_0 = \epsilon, \quad E^{\lambda}_{n+1}= U_0^2(n+r)^2, \quad n=0,1,2,\ldots
\label{enerPT2}
\ee

%%%%%%%%%%%%%%%%%%%%%%
\begin{figure}[htb]
\centering
\subfigure[$\psi^{\lambda}_0(x)$]{\includegraphics[width=0.3\textwidth]{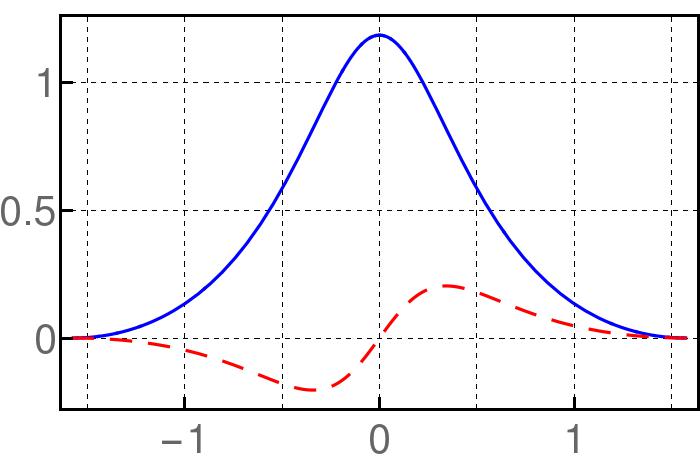}} 
\hspace{1ex}
\subfigure[$\psi^{\lambda}_1(x)$]{\includegraphics[width=0.3\textwidth]{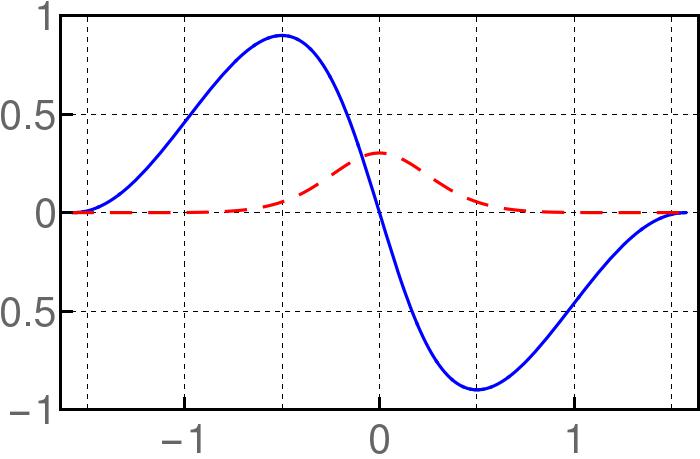}}
\hspace{1ex}
\subfigure[$\psi^{\lambda}_2(x)$]{\includegraphics[width=0.3\textwidth]{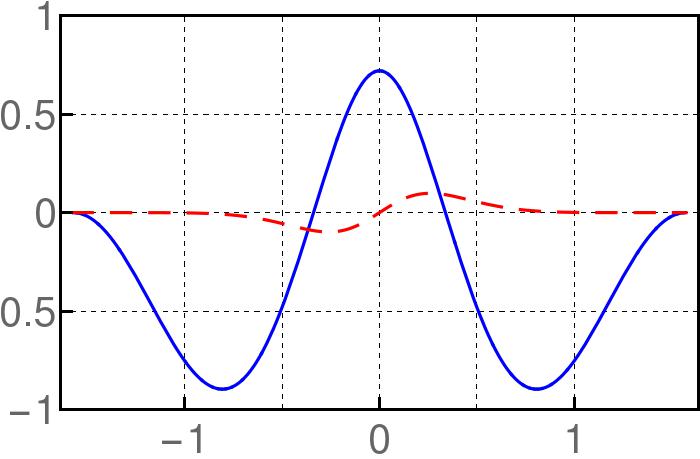}}

\caption{\footnotesize 
(Color online) Real (blue-solid) and imaginary (red-dashed) parts of the first three eigenfunctions belonging to the PT-symmetric potential ($r=3$) depicted in Fig.~\ref{FigPT3}.}
\label{FigPT4}
\end{figure}
%%%%%%%%%%%%%%%%%%%%%%

As regards the complex-valued deformations of trigonometric P\"oschl-Teller potential (\ref{PT}), they can be constructed to be either invariant or not invariant under PT-transformations since $V_0(x)$ is even. The former case is shown in Fig.~\ref{FigPT3} for two different values of the parameter $r$. In both cases $\mbox{Re}V_{\lambda}(x)$ is even, with a very localized symmetric deformation that permits the presence of the new energy $\epsilon$ (in the figure, $\epsilon=1/4$). In turn, $\mbox{Im}V_{\lambda}(x)$ is odd and  satisfies the condition of zero total area (\ref{zero}). That is, the spatial reflection of $\mbox{Im}V_{\lambda}(x)$ is compensated by the complex conjugation ($i \rightarrow -i$). The first three eigenfunctions of such potential are depicted in Figure~\ref{FigPT4} for $r=3$. Notice that they satisfy the interlacing theorems indicated in \cite{Jai17}. 

%%%%%%%%%%%%%
\begin{figure}[htb]
\centering
\subfigure[$r=3$]{\includegraphics[width=0.3\textwidth]{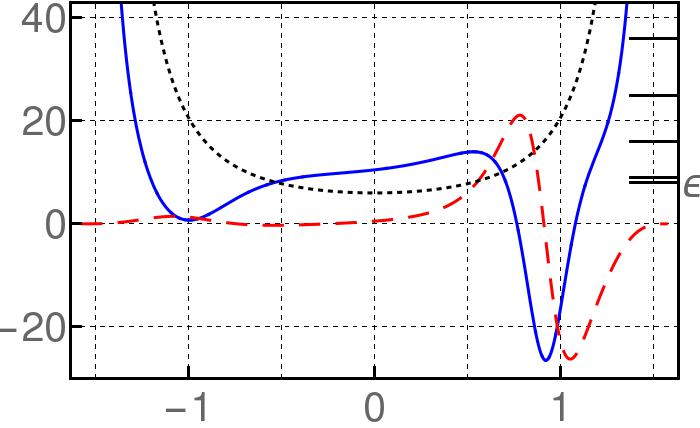}} 
\hspace{5ex}
\subfigure[$r=4$]{\includegraphics[width=0.3\textwidth]{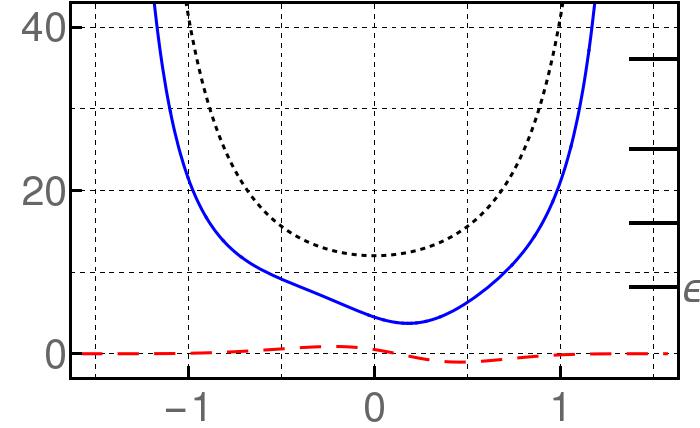}}

\caption{\footnotesize 
(Color online) Real (blue-solid) and imaginary (red-dashed) parts of the complex-valued potentials $V_{\lambda}(x)$ associated with those of Fig.~\ref{FigPT1} (black-dotted). The parameters are $J=1.34$, $I_0=-2.13$, and $\lambda=\sqrt{1.34}$. The ticks on the right-vertical scale refer to energies (\ref{enerPT2}), with $\epsilon=8.075$. 
}
\label{FigPT5}

\end{figure}
%%%%%%%%%%%

On the other hand, Figure~\ref{FigPT5} includes two examples of complex-valued P\"oschl-Teller potentials $V_{\lambda}(x)$ that are not invariant under PT-transformations. In this case the deformation of $\mbox{Re} V_{\lambda}(x)$ is asymmetrical, although this continues to host the new energy level $\epsilon$ (in the figure, $\epsilon = 8.075$). In addition, $\mbox{Im} V_{\lambda}(x)$ still satisfies the condition of zero total area (\ref{zero}) but it is also asymmetrical.

%%%%%%%%%%%%%%%%%%%%%%
\begin{figure}[htb]
\centering
\subfigure[$\psi^{\lambda}_0(x)$]{\includegraphics[width=0.3\textwidth]{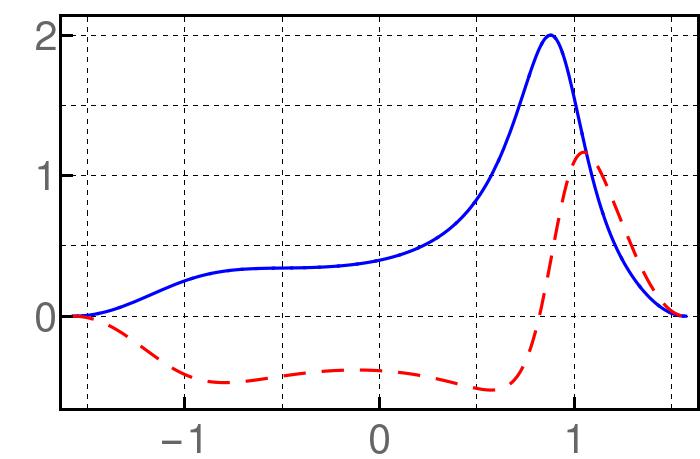}} 
\hspace{1ex}
\subfigure[$\psi^{\lambda}_1(x)$]{\includegraphics[width=0.3\textwidth]{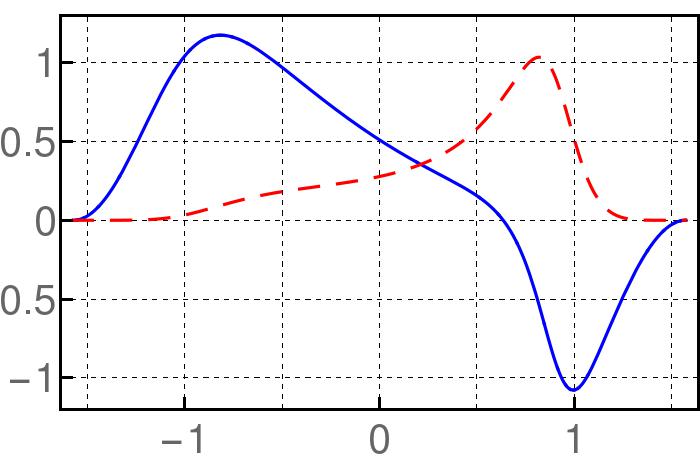}}
\hspace{1ex}
\subfigure[$\psi^{\lambda}_2(x)$]{\includegraphics[width=0.3\textwidth]{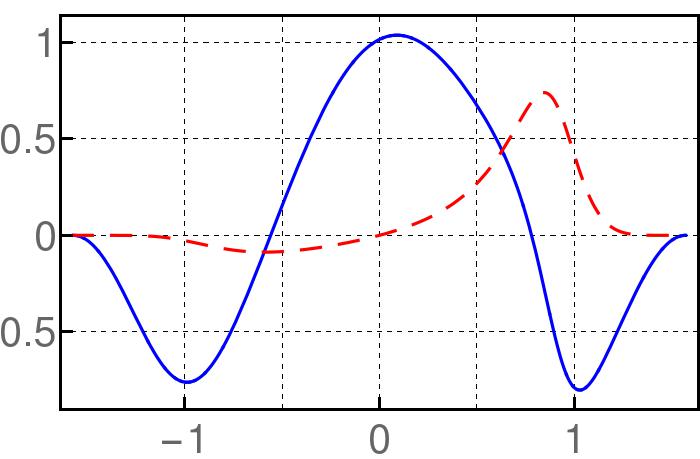}}

\caption{\footnotesize 
(Color online)  Real (blue-solid) and imaginary (red-dashed) parts of the first three eigenfunctions of  the complex-valued potential ($r=3$) depicted in Fig.~\ref{FigPT5}.  
}
\label{FigPT6}
\end{figure}
%%%%%%%%%%%%%%%%%%%%%%

The first three eigenfunctions of these non PT-symmetric potentials are depicted in Fig.~\ref{FigPT6} for $r=3$ and the parameters of Fig.~\ref{FigPT5}. We can appreciate that $\mbox{Re} \, \psi_{n+1}$ has a zero between two zeros of $\mbox{Im} \, \psi_{n+1}$, so the theorems presented in \cite{Jai17} are satisfied. 

%---------------------------------------> Section
\section{The quest of new models}
\label{quest}

The approach presented in sections~\ref{E-S} and \ref{properties} has been addressed to get complex-valued potentials $V_{\lambda}(x)$ with real energy spectrum, no matter if they are PT-symmetric or not. In such a trend it is necessary to take $\lambda \neq 0$. However, the method is general enough to include conventional transformations and produce new integrable potentials of real value as well. Indeed, as indicated above, $\lambda=0$ brings $\beta_{\lambda=0}$ to its conventional form (\ref{beta}) since $\mbox{Im}\beta_{\lambda} =\lambda/\alpha^2$ becomes zero in (\ref{betac}). Remarkably, even in this case the $\alpha$-function is defined by (\ref{alpha}), with $c$ reduced to $c= I_0^2/J$ in (\ref{const}). In turn, the relationship between $a$, $b$ and $c$ becomes simpler $b = \pm 2\sqrt{ac}$, so $\alpha$ is reduced to the linear superposition
\be
\alpha (x) = \sqrt{a} v(x) \pm \sqrt{c} u_p(x) = \sqrt{a \omega_0^2} u_p(x) \left[ \int^x u_p^{-2}(y) dy \pm \frac{I_0}{J}
\right].
\label{alpha2}
\ee

%%%%%%%%%%%%%%%%%%%%%%
\begin{figure}[htb]
\centering
\subfigure[$\gamma_M=1.35$]{\includegraphics[width=0.3\textwidth]{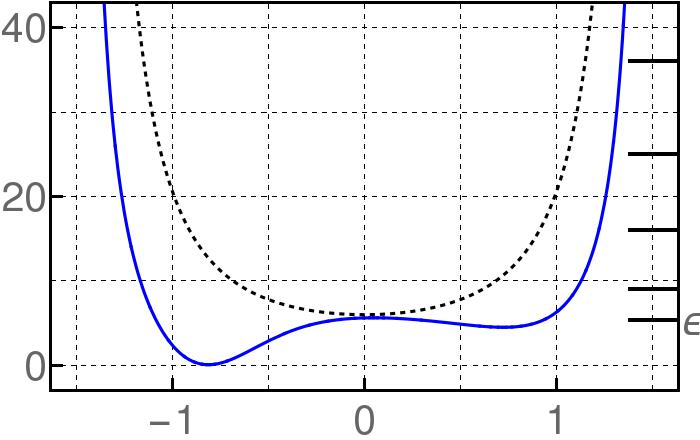}} 
\hspace{5ex}
\subfigure[$\gamma_M=0.74$]{\includegraphics[width=0.3\textwidth]{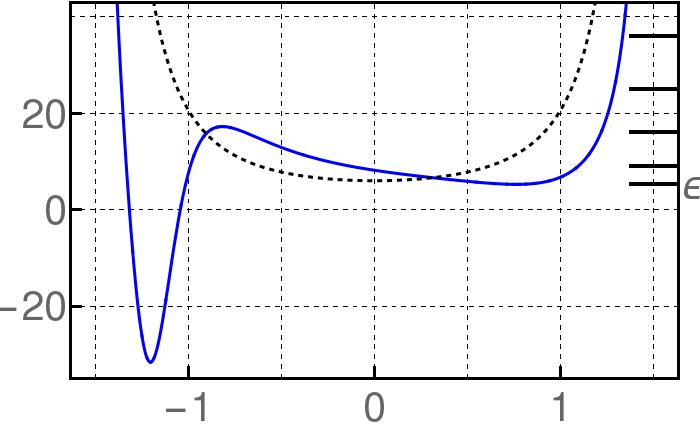}}

\caption{\footnotesize (Color online)  Real-valued Darboux deformations (blue-solid) of the trigonometric P\"oschl-Teller potential (black-dotted) for $r=3$ and $U_0=1$, with $\lambda=0$, $\epsilon=5.26$. The indicated values of $\gamma_M = I_0/J$ correspond to $I_0=3.701$, with $J=2.74$ (a) and $J=5$ (b).
}
\label{FigQuest1}
\end{figure}
%%%%%%%%%%%%%%%%%%%%%%

The last result verifies that our approach includes the one-step Darboux deformations introduced in \cite{Mie84} to construct one-parameter families of solvable real-valued potentials. That is, (\ref{alpha2}) gives rise to the general real-valued solution of the Riccati equation (\ref{riccati1}), where the quotient $\pm I_0/J$ corresponds to the integration constant associated with two successive quadratures \cite{Inc56}. In the Milenik's approach \cite{Mie84}, the quantity $\gamma_M =\vert I_0/J \vert$ is defined such that (\ref{alpha2}) is free of zeros in $\mbox{Dom} V_0$, and parameterizes the family of real-valued potentials $V_{\lambda=0}(x; \gamma_M)$ that is constructed from $V_0(x)$ \cite{Mie04,Boy98,Ros99}. As immediate example we show in Fig.~\ref{FigQuest1} a pair of real-valued potentials $V_{\lambda=0}(x; \gamma_M)$, constructed from the trigonometric P\"oschl-Teller system (\ref{PT}). The spectrum of these potentials is still given by (\ref{enerPT2}) and their eigenfunctions are also constructed from (\ref{even})-(\ref{odd}) with the help of (\ref{darboux}).

On the other hand, our method permits also the construction of complex-valued potentials $V_{\lambda}(x)$ for which the `new' energy $\epsilon$ is added at any position of the discrete spectrum of $V_0(x)$. To be precise, in conventional one-step approaches (Darboux transformations, intertwining techniques, factorization method, Susy QM, etc) the seed function $u$ satisfies (\ref{u}) with $\epsilon \leq E_0$, where $E_0$ is either the ground state energy or the low bound of the continuum spectrum of $V_0(x)$ \cite{Sam99}. However, if $E_n < \epsilon \leq E_{n+1}$, the seed function $u$ has nodes in $\mbox{Dom}V_0$ that produce singularities in the new potential $V(x)=V_0(x) + 2\beta'$. The problem has been circumvented by introducing irreducible second-order (two-step Darboux) transformations where two different energies fulfilling $E_n \leq \epsilon_1 < \epsilon_2 \leq E_{n+1}$ are added \cite{Sam99}. Yet, the limit $\epsilon_2 \rightarrow \epsilon_1$ permits to avoid the problem in elegant form and gives rise to the confluent version of Susy QM \cite{Mie00}. Additional results on two-step Darboux transformations can be found in e.g. \cite{Ber10a,Ber10b,Mid11}. Nevertheless, staying in the first order approach, the oscillation theorems satisfied by the seed function $u$ with eigenvalue $E_n \leq \epsilon \leq E_{n+1}$ prohibit the construction of real-valued potentials $V(x)$ that are free of singularities in $\mbox{Dom}V_0$ \cite{Sam99}. The situation is different for the complex-valued potentials $V_{\lambda}(x)$ since the nonlinear superposition of $u_p$ and $v$ removes the possibility of zeros in (\ref{alpha}), so the function (\ref{newpot}) is regular on $\mbox{Dom}V_0 \subset \mathbb R$.
 
 %%%%%%%%%%%%%
\begin{figure}[htb]
\centering
\subfigure[$\epsilon= \tfrac{E_1+ E_0}{2}$]{\includegraphics[width=0.4\textwidth]{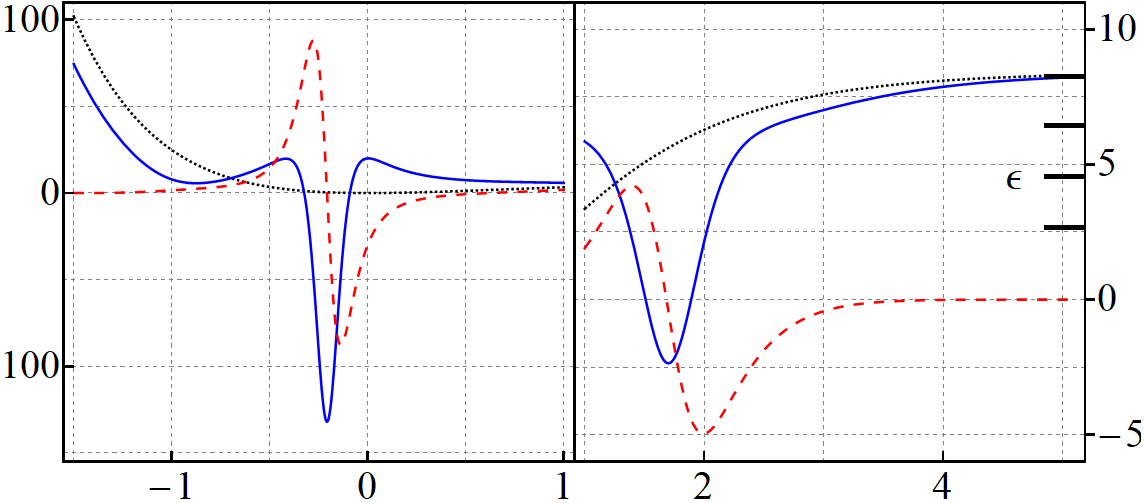}} 
\hspace{1ex}
\subfigure[$\epsilon = \tfrac{E_1+E_0}{2},  \lambda=0$]{\includegraphics[width=0.4\textwidth]{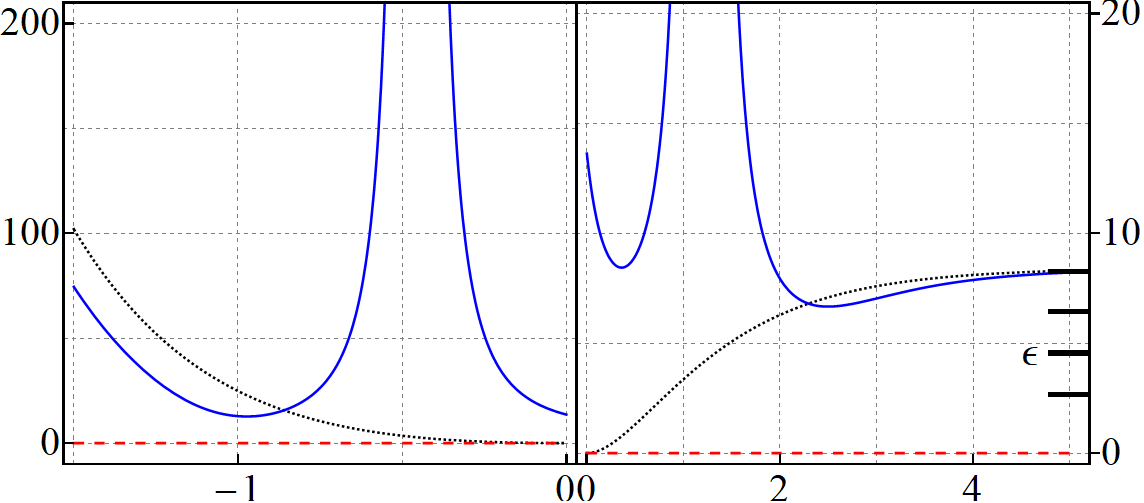}} 
\vskip1ex
\subfigure[$\epsilon= E_1$]{\includegraphics[width=0.4\textwidth]{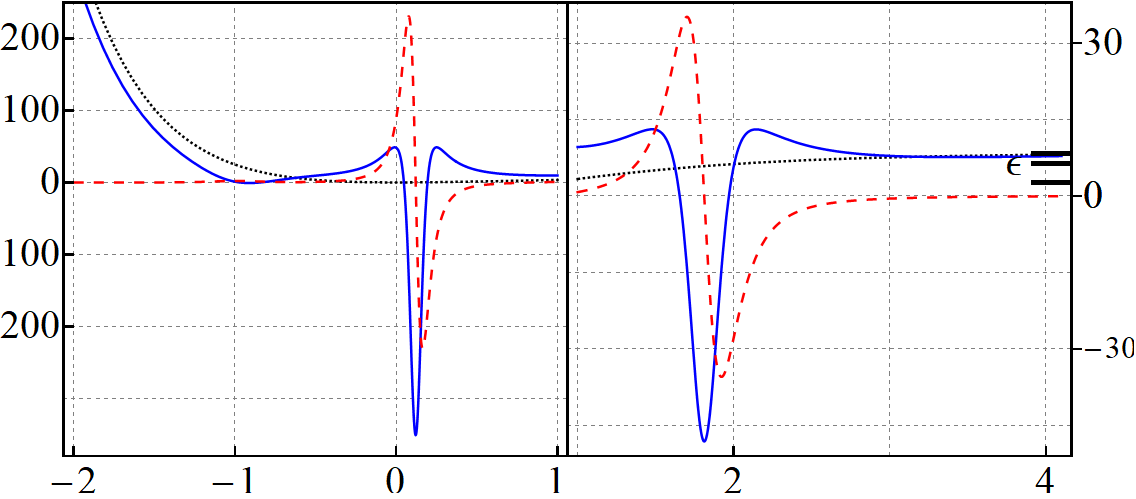}} 
\hspace{1ex}
\subfigure[$\epsilon= E_1,  \lambda=0$]{\includegraphics[width=0.4\textwidth]{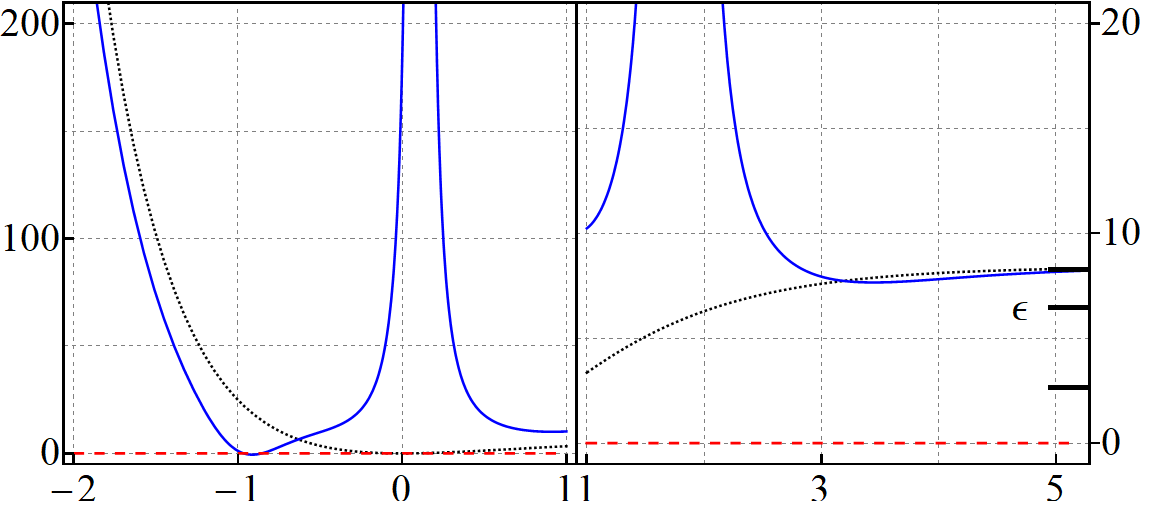}}

\caption{\footnotesize 
(Color online) Potentials generated from the Morse system (\ref{morse1}) by embedding the `new' energy $\epsilon$ in the interval $E_0 \leq \epsilon \leq E_N$, defined by the $N+1$ discrete energies (\ref{morsen}) of bound states (\ref{solMc1}), at the indicated positions. In all cases $\gamma=1$, $\delta=0.4$, $N=2$ and $J=I_0=1$. The left column includes complex-valued potentials with $\lambda =1$ and the right one shows the real valued potentials with $\lambda=0$.
}
\label{FigQuest2}

\end{figure}
%%%%%%%%%%%

As example consider the potentials shown in Fig.~\ref{FigQuest2}. In all cases we have used the Morse system (\ref{morse1}) with $N=2$ as the initial potential $V_0(x)$. The upper row includes potentials generated with $\epsilon = \tfrac{E_1 + E_0}{2}$ and either $\lambda \neq 0$ or $\lambda=0$, Figs.~\ref{FigQuest2}(a) and \ref{FigQuest2}(b) respectively. In both cases the energy spectrum is given by $E_0^{\lambda} = E_0$, $E_2^{\lambda} = E_1$, $E_3^{\lambda} = E_2$, plus the new eigenvalue $E_1^{\lambda}=\epsilon$ that is located between the ground ($E_0$) and first excited ($E_1$) energies of the initial system. The real-valued potential $V_{\lambda=0}(x; \gamma_M)$ shown in Fig.~\ref{FigQuest2}(b) is singular at two different points of $\mbox{Dom}V_0$. This is a consequence of the oscillation theorem obeyed by the general solution (\ref{alpha2}) of Eq.~(\ref{u}) with $E_0 \leq \epsilon \leq E_1$. In Susy QM one says that such a potential is ill defined since $\mbox{Dom}V_0$ is not preserved by the Darboux transformation of $V_0(x)$ \cite{Mar98}. Clearly, this is not the case of the complex-valued potential $V_{\lambda}(x)$ depicted in Fig.~\ref{FigQuest2}(a) since it is regular on $\mathbb R$. In this case, the nonlinear superposition (\ref{alpha}) regularizes the behavior of $\mbox{Re} V_{\lambda}(x)$ and $\mbox{Im} V_{\lambda}(x)$. Besides, at the points where $V_{\lambda=0}(x; \gamma_M)$ is singular, the $\alpha$--function (\ref{alpha}) produces only local deformations in $\mbox{Re} V_{\lambda}(x)$ and $\mbox{Im} V_{\lambda}(x)$.

A similar situation holds for the potentials depicted on the lower row of Fig.~\ref{FigQuest2}. There the `new' energy $\epsilon$ is located exactly at the position of the first excited energy of $V_0(x)$, so the new potentials are exactly isospectral to the Morse system (\ref{morse1}) with $N=2$. The real-valued potential $V_{\lambda=0}(x; \gamma_M)$ of Fig.~\ref{FigQuest2}(d) is ill defined since it has two singularities in $\mbox{Dom}V_0$. In contrast, the complex-valued potential shown in Fig.~\ref{FigQuest2}(c) is regular on $\mathbb R$ with local variations of its real and imaginary parts.

The above examples show that our method produces results that are not available in the conventional one-step Darboux approaches. The complex-valued potentials $V_{\lambda}(x)$ are regular even if $\epsilon$ is embedded in the discrete spectrum of $V_0(x)$, located at arbitrary positions of $E_n \leq \epsilon \leq E_{n+1}$ for any $n=0,1,\ldots$ However, some caution is necessary since the state $\psi_{\epsilon}(x)$, although normalizable in the conventional form, may be of zero binorm \cite{Ros15}. The latter is neither accidental nor rare in physics \cite{Nar03}, so it deserves special attention \cite{Sok06} and will be discussed in detail elsewhere.

%---------------------------------------> Section
\section{Summary and outlook}

We demonstrated that the combination of nonlinear complex-Riccati and Ermakov equations brings out some subtleties of the Darboux theory that are hidden in the conventional studies of integrable models in quantum mechanics. Within this generalized Darboux approach, we have constructed complex-valued potentials that represent non-Hermitian systems with real energy spectrum, no matter if they are PT-symmetric or not. We provided new systems with the discrete energy spectrum of either Morse or trigonometric P\"oschl-Teller potentials as concrete examples. The conventional one-step Darboux approach, giving rise to new solvable Hermitian models, is easily recovered from that introduced here by the proper choice of parameters.

A striking feature of our method is the possibility of adding a real eigenvalue $\epsilon$ that may coincide with the energy $E_n$ of any excited bound state of $V_0(x)$ without producing singularities in the new complex-valued potential $V_{\lambda}(x)$. Remarkably, this is not possible in conventional one-step Darboux approaches since the oscillation theorem prevent the use of discrete energies, other than $E_0$, to produce new potentials with no singularities. In this context, it is to be expected that our method can be applied to study the emergence of exceptional points \cite{Kat66} in the scattering energies of complex-valued potentials.

Other applications may include the study of electromagnetic signals propagating in waveguides, where the Helmholtz equation is formally paired with the Schr\"odinger one \cite{Cruz15a,Cruz15b}. In such a picture the complex-valued potential $V_{\lambda}(x)$ can be identified with a refractive index of balanced gain/loss profile \cite{Ele17}. The study of non-Hermitian coherent states associated with finite-dimensional systems \cite{Gue18} is also available. Finally, the approach can be extended either by applying conventional one-step Darboux transformations on the complex-valued potential $V_{\lambda}(x)$ or by iterating the procedure presented in this work. Further insights may be achieved from the Arnold and point transformations \cite{Gue13,Gue15,Gun17,Car18}. Results in these directions will be reported elsewhere.

%---------------------------------------> Section
\section*{Acknowledgments}

We acknowledge the financial support from the Spanish MINECO (Project MTM2014-57129-C2-1-P) and Junta de Castilla y Le\'on (VA057U16). KZ and ZBG gratefully acknowledge the funding received through the CONACyT Scholarships 45454 and 489856, respectively.

%---------------------------------------> Bibliography

\end{document}